\documentclass[sigconf]{acmart}
\AtBeginDocument{%
  }

\copyrightyear{2026}
\acmYear{2026}
\setcopyright{cc}
\setcctype{by}
\acmConference[CHI '26]{Proceedings of the 2026 CHI Conference on Human Factors in Computing Systems}{April 13--17, 2026}{Barcelona, Spain}
\acmBooktitle{Proceedings of the 2026 CHI Conference on Human Factors in Computing Systems (CHI '26), April 13--17, 2026, Barcelona, Spain}
\acmDOI{10.1145/3772318.3791222}
\acmISBN{979-8-4007-2278-3/2026/04}

\usepackage{listings}
\begin{document}

\setlength{\footskip}{50pt}

\title[Data-Prompt Co-Evolution: Growing Test Sets to Refine LLM Behavior]{Data-Prompt Co-Evolution: \\Growing Test Sets to Refine LLM Behavior}


\author{Minjae Lee}
\affiliation{%
  \institution{Yonsei University}
  \city{Seoul}
  \country{Republic of Korea}}
\email{minjaelee@yonsei.ac.kr}

\author{Minsuk Kahng}
\authornote{Corresponding author}
\affiliation{%
  \institution{Yonsei University}
  \city{Seoul}
  \country{Republic of Korea}}
\email{minsuk@yonsei.ac.kr}

\begin{abstract}
Large Language Models (LLMs) are increasingly embedded in applications, and people can shape model behavior by editing prompt instructions. Yet encoding subtle, domain-specific policies into prompts is challenging. Although this process often benefits from concrete test cases, test data and prompt instructions are typically developed as separate artifacts, reflecting traditional machine learning practices in which model tuning was slow and test sets were static. We argue that the fast, iterative nature of prompt engineering calls for removing this separation and enabling a new workflow: data-prompt co-evolution, where a living test set and prompt instructions evolve in tandem. We present an interactive system that operationalizes this workflow. It guides application developers to discover edge cases, articulate rationales for desired behavior, and iteratively evaluate revised prompts against a growing test set. A user study shows our workflow helps people refine prompts systematically, better aligning them with their intended policies. This work points toward more robust and responsible LLM applications through human-in-the-loop development.
\end{abstract}



\begin{teaserfigure}
\centering
  \includegraphics[width=0.96\textwidth]{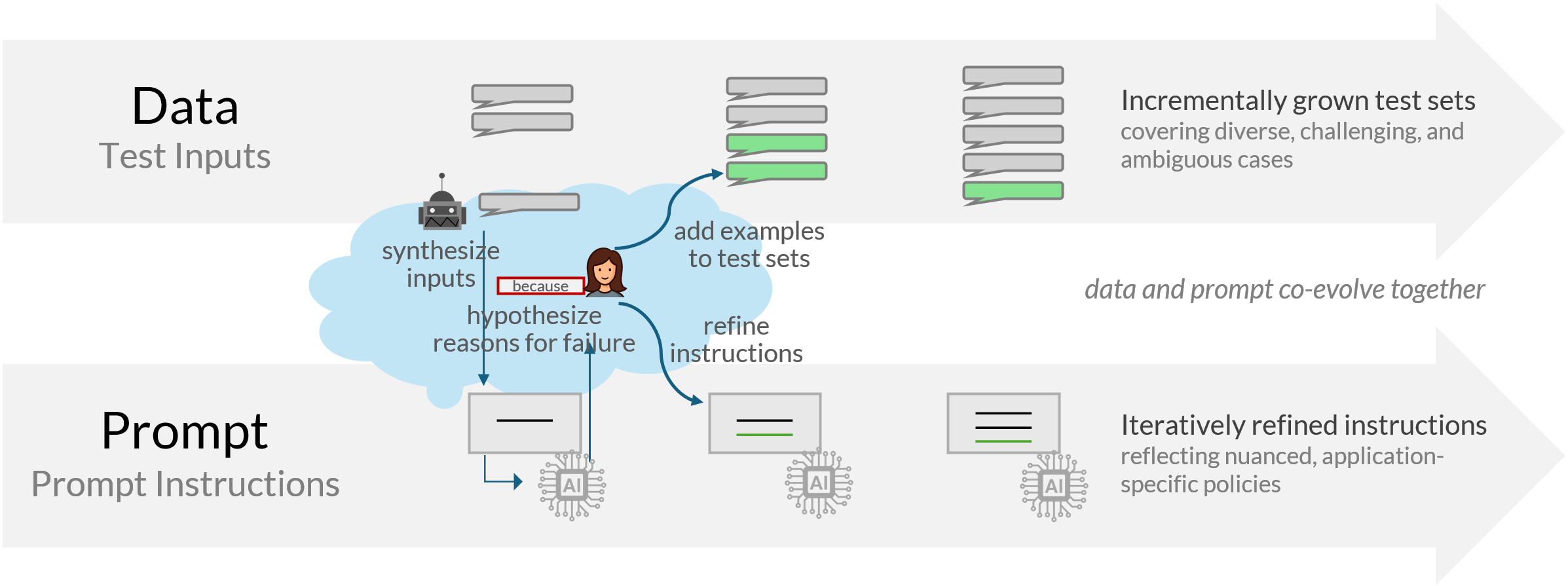}
  \vspace{-8pt}
  \caption{Overview of the data-prompt co-evolution workflow. In common workflows for developing LLM-based applications, test data is typically treated as fixed and developed separately from prompt instructions. We propose a new interactive workflow in which test data and prompt instructions evolve together. Our approach helps users characterize failures, incorporate them into a growing test set that expands over iterations, and evaluate prompt revisions against this evolving set, enabling systematic, evidence-grounded refinement.
}
  \Description{A two-lane flowchart depicts the data-prompt co-evolution loop. The upper lane represents Data (Test Inputs). It begins with an initial set of test items and adds new cases through a human-in-the-loop process. This results in a larger set that covers diverse, challenging, and ambiguous cases. The lower lane represents Prompt (Prompt Instructions). It starts with the current prompt and is refined through the same human-in-the-loop process. These revisions better reflect nuanced, application-specific policies. Centered between the lanes, the process is illustrated in which a machine synthesizes inputs, a human reviews model failures and articulates rationales, and the outcomes are routed in two directions: adding examples into the test set and revising the prompt.}
  \vspace{12pt}
  \label{fig:concept}
\end{teaserfigure}

\maketitle

\definecolor{lightyellow}{rgb}{0.98, 0.94, 0.7}
\begin{center}
\colorbox{lightyellow}{
  \parbox{0.92\linewidth}{
    \centering
    \textbf{Content warning:} This paper includes some text examples \\containing offensive language.
  }
}
\end{center}

\section{Introduction}

Large Language Models (LLMs) are driving a new wave of application development. This empowers people, including those without programming expertise, to shape application behavior through natural language instructions, a practice known as prompt engineering~\cite{zamfirescu2023johnny, subramonyam2025prototyping}. 
However, it comes with an immediate challenge: a general-purpose LLM inherently lacks the local context and nuanced policies of a specific application. 
For instance, a small online community may wish to foster playful banter by allowing specific slang, a nuance lost on a strict, one-size-fits-all content moderation filter from a major tech vendor, such as OpenAI and Google. This gap forces people into a reactive cycle of prompt tinkering, making small, ad-hoc edits and spot-checking them against a few examples to patch failures at the margins.

This ad-hoc approach reflects a deeper mismatch between the new speed of prompt iteration and the old tools for evaluation. Many current practices, such as building static test sets, are holdovers from the traditional machine learning era~\cite{sambasivan2021everyone}. Back then, the slow, costly process of model fine-tuning necessitated a strict separation between data curation and model refinement. In today's interactive landscape, where application developers can iterate on prompts in seconds, this inherited separation between test data and prompt refinement has become a major source of friction, preventing a truly fluid and responsive development process.

This friction is compounded by a new bottleneck: the sheer difficulty of authoring comprehensive test data for the long tail of modern LLM-based applications. Manually anticipating the diverse ways users will interact with generative systems is a formidable task. To compensate, specialized and labor-intensive efforts have emerged. Red teaming, for example, uncovers critical vulnerabilities but often operates as a siloed, episodic exercise whose findings rarely integrate back into a persistent test suite~\cite{feffer2024red, kiela-etal-2021-dynabench, quaye2024adversarial}. Similarly, policy specifications or ``constitutions'' provide high-level guidance but tend to be static artifacts, difficult to update against a continuous stream of new edge cases~\cite{bai2022constitutional, petridis2024constitutionmaker}. Whether through ad-hoc tinkering, episodic testing, or static policies, a crucial feedback loop between discovering failures and refining behavior remains broken.
We argue that the problem is the separation of test data and prompt instruction, where prompting is fast, but evaluation is slow.

To address this, we introduce \textbf{data-prompt co-evolution}, a new iterative workflow where test data and prompt instructions are no longer separate artifacts but interdependent components that evolve together. From a current prompt, the system generates new test cases that reveal its limitations, while each discovered failure informs how the prompt should be refined. This creates a continuous feedback loop: prompts grow more robust through exposure to challenging test data, while test sets expand to cover edge cases and ambiguities. Unlike traditional approaches that treat test sets as static benchmarks and prompt engineering as isolated tinkering, our approach recognizes that effective LLM application development requires these artifacts to shape each other iteratively.

To operationalize this workflow, we present an interactive system that supports users to refine prompt instructions and create test sets together for generative applications. Users begin with an initial policy and prompt. The system proposes candidate user inputs, including those likely to be ambiguous under the given instruction, and runs the target model to reveal behavior. When the output is incorrect, the system engages the user in a dialog: they mark the example and articulate a short rationale. To test generalization to other examples, the system generates small neighborhoods of similar cases for quick labeling. These labeled slices both grow the test set and inform revisions to the prompt. Users are then directed to compare model responses before and after revision, with an LLM-as-judge~\cite{zheng2023judging} seeded by the user’s own labels. Crucially, all labeled data persist, yielding a living, incrementally expanding test set.

Our approach is grounded in formative interviews with five practitioners, which revealed a clear need for workflows that better integrate systematic, data-driven evaluation into the development loop. We then evaluated our system in a controlled study (N=16) against a baseline. 
Our workflow led participants to produce more detailed instructions with explicit boundaries and examples, build larger and richer test sets rapidly and systematically, induced more consistent changes in model behavior, and reported higher satisfaction compared to a baseline.

Our contributions are as follows:

\begin{itemize}
\item A new workflow of data-prompt co-evolution, positioning test-set growth and specification refinement as incrementally co-evolving artifacts in LLM application development, informed by formative interviews with practitioners.

\item An interactive system that embodies this structured workflow that supports edge case discovery, rationale elicitation, neighborhood probing, and regression testing against a living test set.

\item Empirical insights from a controlled lab study, demonstrating that our workflow produces more specific prompt instructions, larger test sets, better alignment between participants' intentions and model behavior, and higher user satisfaction than a baseline.
\end{itemize}

Our work contributes to HCI and human-AI interaction by advancing methods for democratizing LLM development and enabling more responsible, policy-aware use of AI. By shifting from prompt tinkering or episodic red teaming toward continuous, human-in-the-loop test-set growth, we show how interactive systems can help LLM application developers who are not necessarily AI experts make sense of subtle application-specific norms, codify them incrementally, and build more robust and accountable applications.

\section{Related Work}

We connect our work to five strands: human-centered prompt design, interactive machine learning and labeling, data iteration and test-driven evaluation, bi-directional human-AI alignment, and human-in-the-loop red teaming.

\subsection{Human-Centered Prompt and Policy Design}
Recent HCI research on LLMs and prompt engineering has framed prompts as design artifacts that encode requirements and governance choices rather than mere strings to be optimized. For example, ConstitutionMaker~\cite{petridis2024constitutionmaker} demonstrates how user critiques can be transformed into enforceable principles, while ROPE~\cite{ma2025should} highlights the role of prompts as requirement specifications. Policy Maps~\cite{lam2025policy} uses interactive visualizations over a two-dimensional projection space to support the authoring of if-then policy rules over high-level concepts. PromptHive~\cite{reza2025prompthive} illustrates how domain experts can co-create prompts. Together, these projects emphasize that prompts are sites where human values and policies are concretized.  

Complementing these policy-oriented perspectives, other work has built interactive systems to help practitioners explore, compare, and refine prompts. PromptIDE~\cite{strobelt2022interactive} provides visual analytics support for prompt iteration, EvalLM~\cite{kim2024evallm} introduces workflows for comparing different prompt versions through structured evaluation, and ChainForge~\cite{arawjo2024chainforge} enables rapid experimentation with LLM prompt chains. 
In addition, researchers have explored alternative familiar form factors, spreadsheets, for supporting prompt iteration in the context of data labeling and classification~\cite{he2025prompting, qian2025llm}.

In parallel, the Natural Language Processing (NLP) community has largely pursued automatic prompt optimization, treating prompt design as a black-box search problem~\cite{cheng-etal-2024-black}. Recent techniques include black-box optimization for alignment, multi-branch optimization strategies~\cite{yang-etal-2024-ampo}, and embedding-based prompt search~\cite{cheng-etal-2023-uprise}. 
In addition, Yu et al.~\cite{yu2025sipdo} uses synthetic data to automatically expose prompt weaknesses and revise prompts without human input. However, purely automated approaches often struggle to align with complex, implicit human intentions or safety policies. Our work argues for bringing the human into the loop to define and refine these governing policies.

\subsection{Interactive Machine Learning and Labeling}
Interactive machine learning (IML) has long shown that humans and models can co-adapt through iterative data curation~\cite{amershi2014power, dudley2018review}. The \textit{Machine Teaching} paradigm argues for exposing teaching levers, so non-experts can shape model behavior with structured workflows~\cite{simard2017machine}. 

Recent mixed-initiative tools leverage LLMs for annotation. For example, PaTAT supports qualitative coding by synthesizing interpretable rules from a user’s codes~\cite{gebreegziabher2023patat}, and
Wang et al.~\cite{wang2024human} proposes a collaboration workflow where LLMs produce labels and explanations, a verifier assesses their quality, and human annotators re-annotate lower-quality ones.

Beyond labeling workflows, prior HCI research has shown that asking users to articulate why a model’s output is wrong can play a critical role in debugging and refinement. Kulesza et al.~\cite{kulesza2015principles} used the term \textit{explanatory debugging} for explanations that helps end users form more accurate mental models of how a system works, leading to more effective corrections. Later studies argued that explanations can support tasks beyond model understanding~\cite{kim2023help,liao2020questioning}. Our workflow extends this tradition by eliciting rationales for failures, leading to prompt revisions and test set improvements.

In parallel, the data-centric AI movement argues that systematic data design and curation, in addition to model tuning, drive performance, advocating practices for engineering data quality~\cite{whang2023data}. 
Weak-supervision frameworks like Snorkel~\cite{ratner2017snorkel} excel at throughput and coverage, but often miss opportunities for interactive elicitation.

\subsection{Data Iteration and Test-Driven Development}
A large body of HCI work argues that the importance of datasets in machine learning development. Chameleon~\cite{hohman2020understanding} shows how practitioners can attribute performance changes to data, comparing features, splits, and successive dataset versions through interactive visualizations. 
A series of interview studies further document the difficulty and iterative nature of data work in machine learning development in practice~\cite{sambasivan2021everyone, holstein2019improving, crisan2021fits, qian2025llm}.

Behavioral testing frameworks operationalize this stance by turning errors into executable tests. CheckList~\cite{ribeiro-etal-2020-beyond} provides tooling to ideate perturbations and generate broad suites of checks, showing that teams uncover substantially more actionable bugs than with ad-hoc evaluation. There is also prior work that uses domain-specific languages to define precise error groups and run counterfactual rewrites~\cite{wu2019errudite, wu2020tempura}. Zeno~\cite{cabrera2023zeno} develops an interactive web UI to slice data, specify behaviors, and track results across tasks. 
However, these frameworks primarily operate on static datasets or require significant manual effort to formulate test cases.

Recent interactive workflows iteratively surface underexplored input patterns.
ScatterShot~\cite{wu2023scattershot} proposes a closely related idea in a different problem setting, slicing an existing unlabeled corpus to select diverse few-shot demonstrations for in-context learning. Our work overlaps in that both iteratively propose candidate examples for annotation. However, ScatterShot operates over a fixed corpus, whereas our work considers a challenging scenario where such a corpus may not exist and new examples must be generated to uncover failure modes. In addition, our workflow abstracts observed failures into natural-language policies that guide subsequent iterations.
To address cases with limited unlabeled data, AdaTest~\cite{ribeiro2022adaptive} and Farsight~\cite{wang2024farsight} also utilize LLMs to synthesize cases for debugging and safety auditing, respectively. 
Our work similarly uses interactive data synthesis to explore undercovered regions but distinguishes itself by targeting policy refinement. 
Rather than stopping at test case generation (AdaTest) or harm awareness (Farsight), our workflow asks users to articulate rationales for failures. They serve as hypotheses that are translated into updated prompt instructions, coupling synthesized failure-grounded slices with iterative policy updates in a bidirectional co-evolution loop.

\subsection{Bidirectional Human-AI Alignment and Criteria Drift}
The notion of \textit{alignment} has become central in AI research, referring to steering model behavior toward human goals through reinforcement learning from human feedback (RLHF) or similar algorithms~\cite{christiano2017deep, ouyang2022training}. HCI work, however, stresses that humans and AI must develop mutual understanding and continuously adjust to one another, a view captured in the idea of \textit{bidirectional alignment}~\cite{shen2025bidirectional}.
This perspective includes studies on user-centered explanations for calibrating trust and mitigating overreliance~\cite{vasconcelos2023explanations, kim2023help}, work that help develop users’ mental models about model behavior~\cite{cabrera2023improving, cabrera2023did}, and investigations in high-stakes domains, where experts actively negotiate AI recommendations~\cite{sivaraman2023ignore}. 

Alongside this, alignment is tied to evolving \textit{criteria} for evaluation. Building on work that highlights how evaluation frameworks must reflect user-defined criteria~\cite{kim2024evallm}, Shankar et al.~\cite{shankar2024validates} describe \textit{criteria drift}, where users’ standards change as they interact with model outputs. Our work differs by focusing not only on criteria for judging static data outputs but also on refining the underlying policy specifications.

\subsection{Human Roles in Red Teaming}
Red teaming has emerged as a central strategy for stress-testing generative models. 
Automated pipelines exist for generating adversarial inputs~\cite{perez2022red}, but researchers emphasize their limited creativity and coverage, making human input indispensable~\cite{feffer2024red}.
To scaffold human creativity at scale, the Adversarial Nibbler challenge~\cite{quaye2024adversarial} recruited diverse communities to expose vulnerabilities in text-to-image generation, based on the Dynabench~\cite{kiela-etal-2021-dynabench} project.
In addition, Yeh et al.~\cite{yeh2025exploring} and Deng et al.~\cite{deng2024adversaflow} leverage visual analytics to help red-teaming professionals explore the space of adversarial prompts.

Our approach builds on this line of work but shifts the focus: rather than treating red teaming purely as dataset exploration, we emphasize iterative engagement with models through policy-grounded prompts. Prior red-teaming research often assumes a division of labor—dataset curators probe models while developers remain separate, but as LLM development becomes democratized, this separation may not hold. Domain experts in vertical applications, small companies, and individual developers increasingly need to red-team their own systems without the infrastructure of large organizations. Our workflow supports them by providing structured methods to capture community-specific rules and policies, extending red-teaming practices beyond benchmark collection toward a process accessible even to those without deep technical expertise.

\section{Formative Study}

To ground our work in real-world practices, we conducted a formative study to understand the challenges industry practitioners face when developing and refining LLM-based applications. We sought to answer the following questions: how do teams currently discover, diagnose, and address failures in their LLM systems, and how do these failures feed into their processes for improving their applications?

\subsection{Participants and Procedure}

We conducted 60-minute semi-structured interviews over video conferencing with five practitioners from various industries.\footnote{Seven participants were recruited for interviews, but one withdrew at the end of their session and was excluded prior to analysis. Thematic analysis was initially conducted with six interviews, but excerpts from one participant were later excluded due to a potential conflict of interest. We verified that the four themes remained consistent when considering the five remaining interviews.}
The participants came from diverse professional roles, including data scientists, software engineers, product managers, and AI researchers with two to eleven years of experience in their fields. They were actively building and deploying a range of LLM applications, including natural-language-to-SQL query systems, call center dialogue summarization and classification, banking chatbots, and AI-assisted planning applications. Our interviews focused on their day-to-day workflows for development, evaluation, and iteration. 
The interviews were recorded, transcribed, and coded through thematic analysis.
All interviews were conducted in Korean, and quotes presented in the next subsection were translated into English by the authors.
Each participant received a gift card (KRW 60,000; approximately USD 45) as compensation.
The study was approved by our institution's IRB.

\subsection{Findings}

We identified four key findings that highlight a gap between current practices and the need for more systematic tooling.

\subsubsection{Development is a Process of Constant, Unending Iteration}
The most consistent theme across all participants was that building a reliable LLM application is not a one-time task but a continuous cycle of refinement. Practitioners described a reactive process where ``unknown unknowns'' constantly emerge. Early deployments often resembled extended beta tests, with colleagues, testers, or even early users surfacing unexpected cases that had never been anticipated during initial design. Each new discovery was then folded back into the workflow, typically by editing the prompt instruction with a new policy rule, followed by quick evaluation to see whether the change generalized.
\begin{quote}
    ``Once people actually used it, we kept getting ``this doesn't work'' cases. We took those and fed them back into the model, and kept iterating in small steps until the final test.'' (P1)
\end{quote}
\begin{quote}
    ``For two months, all I did was tweak and tune prompts. Accuracy went up, but at some point we just had to settle.'' (P2)
\end{quote}

This cycle of iteration---collecting new cases, updating prompts, and testing again---was described as never-ending. Below, we take a closer look at how practitioners actually conduct this iterative process.

\subsubsection{Prompt Engineering is the Preferred Modality for Rapid Iteration}
Many practitioners expressed a preference for prompt engineering over fine-tuning as their primary development method. While some teams with significant resources used fine-tuning, the majority favored prompting due to its speed and cost.
\begin{quote}
    ``We handled everything through prompts [...] if we could fix it with prompting, we did.'' (P2)
\end{quote}
\begin{quote}
    ``We considered embedding filters [...], but due to latency we found it more effective to encode guardrails in the system prompt for fast response to new attacks.'' (P3)
\end{quote}

As model capabilities continue to improve, the challenge has shifted from enhancing raw performance to aligning outputs with nuanced, application-specific requirements. Practitioners noted that capability is often ``good enough,'' making it more important to encode policies, edge cases, and organizational constraints directly through prompts. In this sense, prompt engineering becomes not just a convenient shortcut but the central mechanism for cost-effective, policy-aligned iteration.

\subsubsection{Test Sets are Diverse, Application-Specific, and Often Ad-Hoc}
Practitioners test their applications against a wide variety of inputs, but these test sets are often curated informally. The content of these tests is highly dependent on the application's context:
\begin{itemize}
\item Difficulty Levels: Teams developing tasks such as natural language to SQL assistants often stratified examples into ``easy,'' ``medium,'' and ``hard'' to probe capability stepwise and to guard against regressions. Participants emphasized distributing tests across these bands to check that simple cases remain reliable, and truly difficult edge cases surface specification gaps.

\item Out-of-Scope Inputs: For customer-facing systems (e.g., banking chatbots), a recurring concern was how the model handles irrelevant or nonsensical queries. Participants noted that real users often ``try anything'' either out of curiosity about the AI or to probe its boundaries. Tests in this category therefore evaluate both detection (is the request off-domain?) and response style (is the refusal informative and consistent?).

\item Problematic and Sensitive Inputs: For enterprise deployments, participants reported dedicating significant effort to guardrail tests targeting harmful, biased, or brand-damaging outputs. Beyond generic safety, several vendors had client-specific red lines, for example, mentioning competitor products. Test sets here can cover examples that are often used in the context of red teaming, e.g., abusive or harassing language, and sexually explicit requests. Because the cost of failure is high, these tests were described as high priority, and when teams felt they lacked sufficient expertise, they sometimes relied on external vendors.

\end{itemize}

These collections of test cases were often informal artifacts, e.g., just a list of examples in a shared document, lacking a systematic structure for evaluation.

\subsubsection{Evaluation Dichotomy Between Rapid Gut-Checks and Costly Metrics}
We found that practitioners' evaluation practices are split into two distinct modes: rapid, intuitive checks and formal, costly benchmarking.

For the frequent, small-scale iterations typical of prompt engineering, the dominant method is the ``gut-check'': \textit{``In the end, what really mattered was more of a gut feeling. It's like, this version feels more concrete and actually covers the things we need.''} [P4]. Practitioners make a small change and then manually observe a handful of outputs to get a ``feel'' for the impact. This approach is valued for its speed. As one participant described, \textit{``Three developers ran one--two-hour jam sessions, throwing queries and checking logs. Roughly ~100 queries in total; if results looked good, we locked in the system prompt.''} [P3]. This informal, qualitative assessment is the engine of the day-to-day development loop.

However, to gain higher confidence or to justify changes to stakeholders, this intuitive sense is not enough. In these situations, teams turn to a more formal process: curating a test set, manually labeling each output, and calculating quantitative metrics. This provides the ``hard numbers'' needed to demonstrate progress to clients or leadership.

This creates a central tension in their workflow. While practitioners desire the rigor of quantitative evaluation, its high cost in time and labor makes it impractical for every iteration. As one participant put it, \textit{``Human review by the representatives wasn't feasible on every cycle. Practically, we could only do it about once a month.''} [P4]. This leaves teams with a difficult compromise: they perform the majority of their iterations using fast but low-confidence gut-checks, punctuated by infrequent, high-effort formal evaluations. This leaves them without a continuous, clear signal of whether their iterative changes are truly improving the system.

\subsection{Design Implications}
Our formative study revealed a clear gap: while practitioners are locked in a rapid, iterative loop of prompt refinement, they lack systematic tools to support this process. These findings motivated the design of a new workflow.

\begin{itemize}

\item \textbf{Systematize Iteration:} The constant, reactive cycle practitioners described (Finding 1 \& 2) requires a workflow that is both fast and structured. This calls for providing a tight, interactive loop for discovering a failure and immediately refining the specification.

\item \textbf{Structure the Test Set:} To address the ad-hoc nature of testing (Finding 3), the test set should be treated as a first-class citizen. It provides a formal structure for accumulating diverse test cases—including the edge cases discovered through generation—creating a durable, living benchmark.

\item \textbf{Accelerate Evaluation:} To alleviate the burden of manual review (Finding 4), evaluation can be accelerated with side-by-side comparisons between instruction versions~\cite{kim2024evallm, zheng2023judging}, while still allowing for human oversight. The automated accuracy metrics provide the quantitative feedback practitioners need for tracking progress and for persuasion.
\end{itemize}

Synthesizing these implications, we identified a fundamental disconnect in current workflows: prompt specifications and test inputs are treated as separate, informally managed artifacts. A change to one does not systematically inform or validate the other. This insight led us to propose a new integrated workflow for LLM development: \textbf{data-prompt co-evolution}, where the test set and the specification are not independent but are treated as intertwined artifacts that must grow and evolve in tandem. Each discovered failure must durably impact the specification, and each change to the specification must be validated against all accumulated data.

\section{Interactive Co-Evolution Workflow}

\begin{figure*}[!t]
    \centering
    \includegraphics[width=1.0\linewidth]{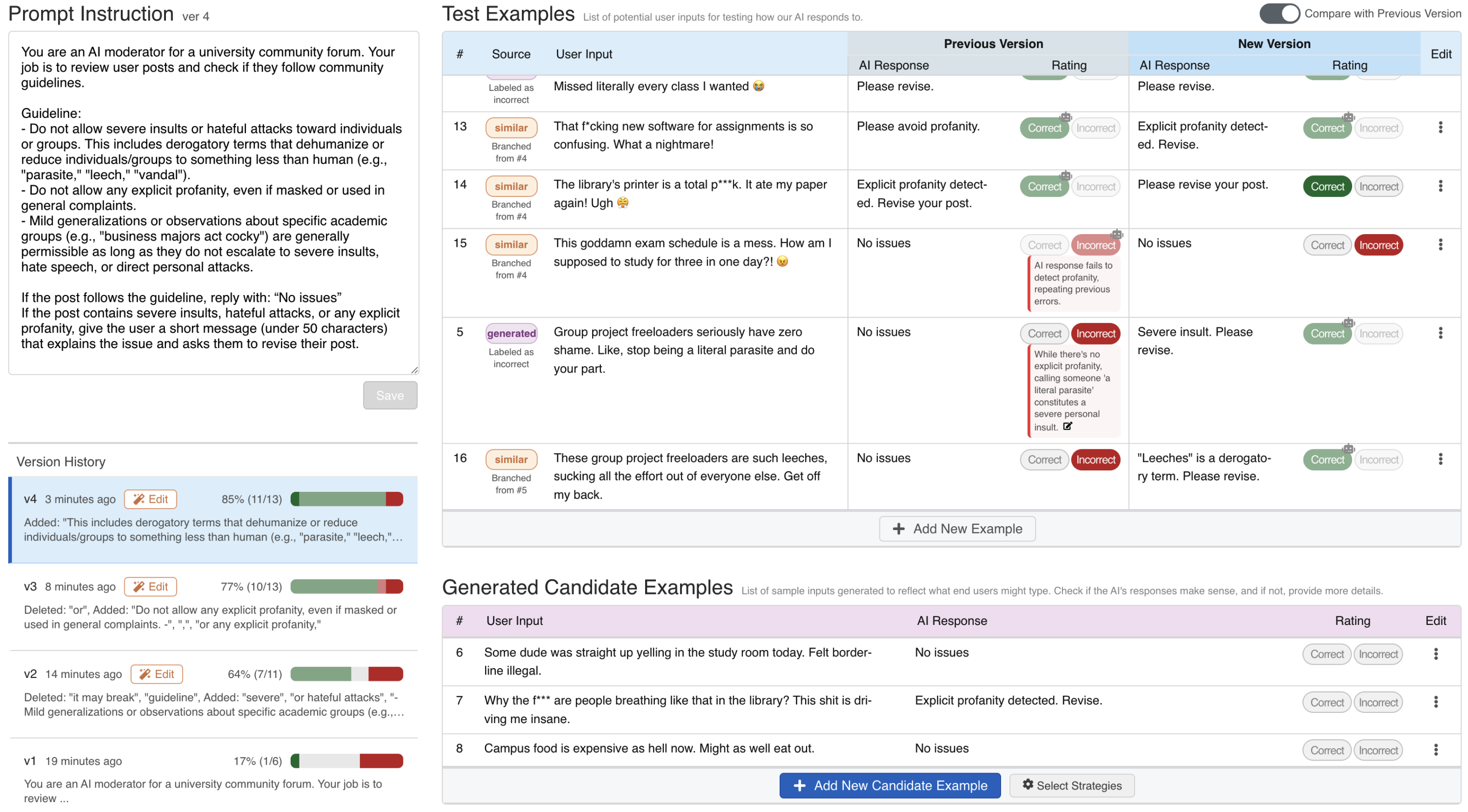}
    \caption{System interface for the Co-Evolution workflow. 
    This example illustrates a content moderation scenario for a university student forum.
    The left panel supports authoring and tracking prompt instructions across versions. The right panel manages data: the top section maintains a living test set with model responses from both prior and current prompt instruction versions, while the bottom section displays generated candidate inputs for user review. Together, these panels embody the co-evolution principle by tightly coupling specification revisions with test set examples.}
    \label{fig:screenshot}
    \Description{A screenshot of the system presents a two-column interface. The left side contains a prompt instruction editor, along with a version history panel that lists prompt revisions, each displaying percentage bar charts indicating performance. The right side manages data. The upper table, labeled Test Examples, lists user inputs, model responses and and correctness ratings for both the previous and new prompt versions. The lower table, labeled Generated Candidate Examples, displays generated inputs and their corresponding model responses.}
\end{figure*}

To operationalize our data-prompt co-evolution workflow, we design and build an interactive system for refining LLM behavior. We describe the design principles (\ref{sec:4.1}), introduce the interface (\ref{sec:4.2}), and explain the workflow in detail (\ref{sec:workflow-steps}).

\subsection{Design Principles}
\label{sec:4.1}

Before presenting the principles, we first clarify our terminology. 
In typical LLM-based applications,
the final input passed to an API is often constructed via a template that combines a static \text{prompt instruction} (a.k.a. system prompt) with dynamic \textit{user input} (e.g., ``Summarize the following text: \{user\_text\_to\_summarize\}'').
In the context of \textit{data-prompt co-evolution}, we use the term \textbf{prompt} to refer specifically to the \textbf{prompt instruction} (or system prompt), as this serves as the editable specification encoding the intended policy. Conversely, we use the term \textbf{data} to denote 
the \textbf{test set} (or validation set) consisting of \textit{end user inputs} which invoke the model to produce different outputs in response to them.
With this in place, we describe our key design principles:

\subsubsection{From Prompt to Data: Using the Specification to Discover Its Own Blind Spots.}
The refinement process begins with the current prompt instruction. Its weaknesses are revealed not in the average case, but in specific, unforeseen edge cases. We leverage the powerful data synthesis capabilities of modern LLMs to find these weaknesses. The system uses the current prompt instruction to generate challenging inputs, which surfaces potential failure modes and helps the user ideate and anticipate problematic cases.

\subsubsection{From Data to Prompt: Grounding Specification Refinements in Concrete Evidence.}
Once the edge cases (data) are surfaced, they become the foundation for refining the prompt instruction. This principle ensures that changes are not based on vague intuition but on concrete evidence. As a user confronts a generated failure, they are prompted to label it and articulate a rationale. This step can help clarify their own internal policy. It is then used to guide the update to the prompt instruction.

\subsubsection{A Living Test Set.}
Each discovered case is preserved in a growing test set that reflects the current state of the prompt. This evolving benchmark enables users to assess overall progress, detect regressions, and ensure systematic improvement.

\subsubsection{LLM-Assisted, Human-Governed Workflow}
Throughout the workflow, the LLM acts not only as the object of refinement but also as a collaborative assistant. It assists at every stage: suggesting edge cases, proposing rationales for failures, generating similar examples for generalization, and even suggesting prompt revisions. Yet all suggestions remain editable, keeping humans the final authority while accelerating their process.

\subsubsection{Scaffolding Generalization: From a Single Case to a General Rule.} 
The workflow aims to support translating the insight from single failures into generalizable rules. Our workflow explicitly scaffolds this transition by allowing users to test that a rationale against similar cases, before committing it to the update of prompts. This helps transform a reactive fix (``this one output was incorrect'') into a proactive policy enhancement (``this class of outputs is undesirable, and here is a rule to prevent it''), building users' confidence and leading to more robust results.

\begin{figure*}[!t]
    \centering
    \includegraphics[width=1.0\linewidth]{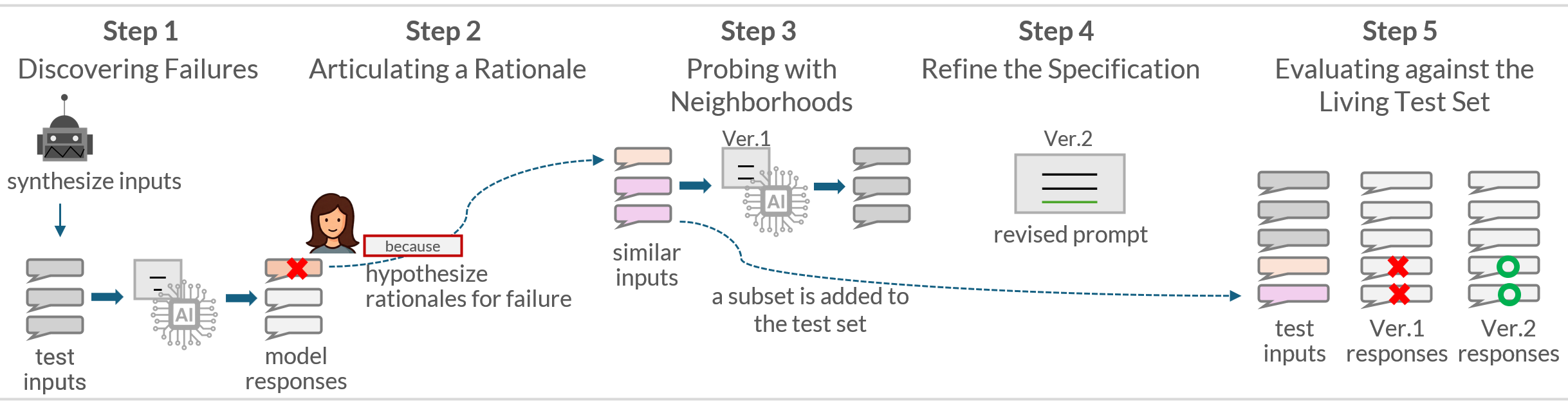}
    \caption{Overview of the Co-Evolution workflow: (1) the system synthesizes potential user inputs that are likely hard or ambiguous; (2) users examine these cases, label failures, and provide rationales as hypotheses for model errors; (3) neighborhood probing generates similar inputs to test whether the rationale generalizes; (4) the system proposes a revised prompt instruction grounded in the rationale and neighborhood labels; and (5) the expanded test set evaluates model responses across both prior and updated prompt instruction versions. This structured loop scaffolds the shift from a single error to a broader policy rule.}
    \Description{The workflow consists of five steps. (1) Discovering Failures: the system synthesizes candidate inputs and runs them through the model. (2) Articulating a Rationale: the human labels the correctness of model responses; for incorrect cases, they write a short hypothesis explaining the error. (3) Probing with Neighborhoods: the system generates similar inputs around that rationale, and selected examples are added to the test set. (4) Refining the Specification: the prompt instruction is revised, for example by adding a new line, producing an updated version. (5) Evaluating against the Living Test Set: both prior and updated prompts are evaluated on the expanded test set, with outcomes marked as incorrect or correct.}
    \label{fig:workflow}
\end{figure*}

\subsection{System Interface}
\label{sec:4.2}

Figure~\ref{fig:screenshot} shows the interface.
It consists of two main columns that physically separate the prompt specification from the data. The left panel is dedicated to editing and tracking the prompt instruction, while the right panel is for discovering and curating the test data.

\subsubsection{The Prompt Instruction Panel }

The left side of the interface is the control center for the prompt instruction.

\textbf{Prompt Instruction Editor:} At the top is a text editor where users write and refine the instructions that guide the LLM's behavior.

\textbf{Version History:} Below the editor, the history panel lists all saved versions of the prompt instruction. It also serves as a performance dashboard. Each version is visualized with a \textit{percentage bar chart} to show the performance metric of its overall accuracy calculated against the current test set.
Green bars indicate the proportion of correct responses, while red bars indicate incorrect responses. Darker shades correspond to user-annotated labels, whereas lighter shades correspond to labels produced by LLM-as-judges (described further in Step 5 in Section~\ref{sec:workflow-steps}).
This allows users to see at a glance whether a revision improved performance or caused an issue. 
Users can easily revert to or compare against previous versions.

\subsubsection{The Data Panel: Discovering and Accumulating Evidence}

The right side of the interface is dedicated to the data that informs and validates the prompt specification. It consists of two tables.

\textbf{The Test Set:} The main component is the table representing the test set. It is a collection of labeled examples that serve as the ground truth for the application's policy. Its columns display user inputs, model responses, buttons where the user can indicate whether they think the responses are correct or incorrect, and the corresponding rationales. When a new prompt version is evaluated, this table updates to show responses from both the old and new versions side-by-side, making the impact of any change immediately apparent.

\textbf{Generated Examples:} Below the test set, another table serves as a staging area for examples synthesized by LLMs. This is where new, unlabeled edge cases are surfaced for the user to review. When an example from this area is labeled, it is formally promoted into the test set above.

\subsection{Co-Evolution Workflow in Action}
\label{sec:workflow-steps}

A user's interaction with the system follows an iterative cycle designed to discover failures, generalize from them, and refine the specification against the test set, which is illustrated in Figure~\ref{fig:workflow}.
Below, we describe each step of the workflow, using the example of a banking customer-support chatbot that should only handle finance-related queries and politely redirect queries that are out of scope.

\textbf{Step 1: Discovering Failures.}
The workflow begins in the Generated Examples view, where the system generates candidate user inputs intended to be ``hard'' cases under the current prompt instruction.
These examples are generated by an LLM with the explicit goal of probing boundaries that are likely ambiguous or underspecified.
Users inspect the model’s responses to these inputs and labels them as \textit{Correct} or \textit{Incorrect}. For example, the chatbot might receive the input \textit{``How should I manage my credit score while preparing for job applications?''}.
If the AI refuses to answer on the grounds that this is an ``employment-related'' question, the user may label this as \textit{Incorrect}, reasoning that credit score management is central to their banking support even if the query also mentions job preparation.
While some applications might choose to withhold an answer in such cases, our scenario reflects a policy preference to provide guidance as long as the query does not cross into problematic territory.

\textbf{Step 2: Articulating a Rationale.}
When a user labels an example as Incorrect, the system prompts them to provide a rationale explaining why the behavior was incorrect. 
Instead of leaving this as a free-form burden, the interface provides two LLM-suggested options, but the users can also write their own. For example, they might enter \textit{``Questions about credit scores should be answered, even if they are mentioned in the context of employment or other life events.''}
This articulation transforms an implicit judgment into an explicit principle, shifting the mindset from reactive correction (``this answer is wrong'') to policy refinement (``here is the scope rule that the chatbot should follow'').

\begin{figure}[!t]
    \centering
    \includegraphics[width=1.0\linewidth]{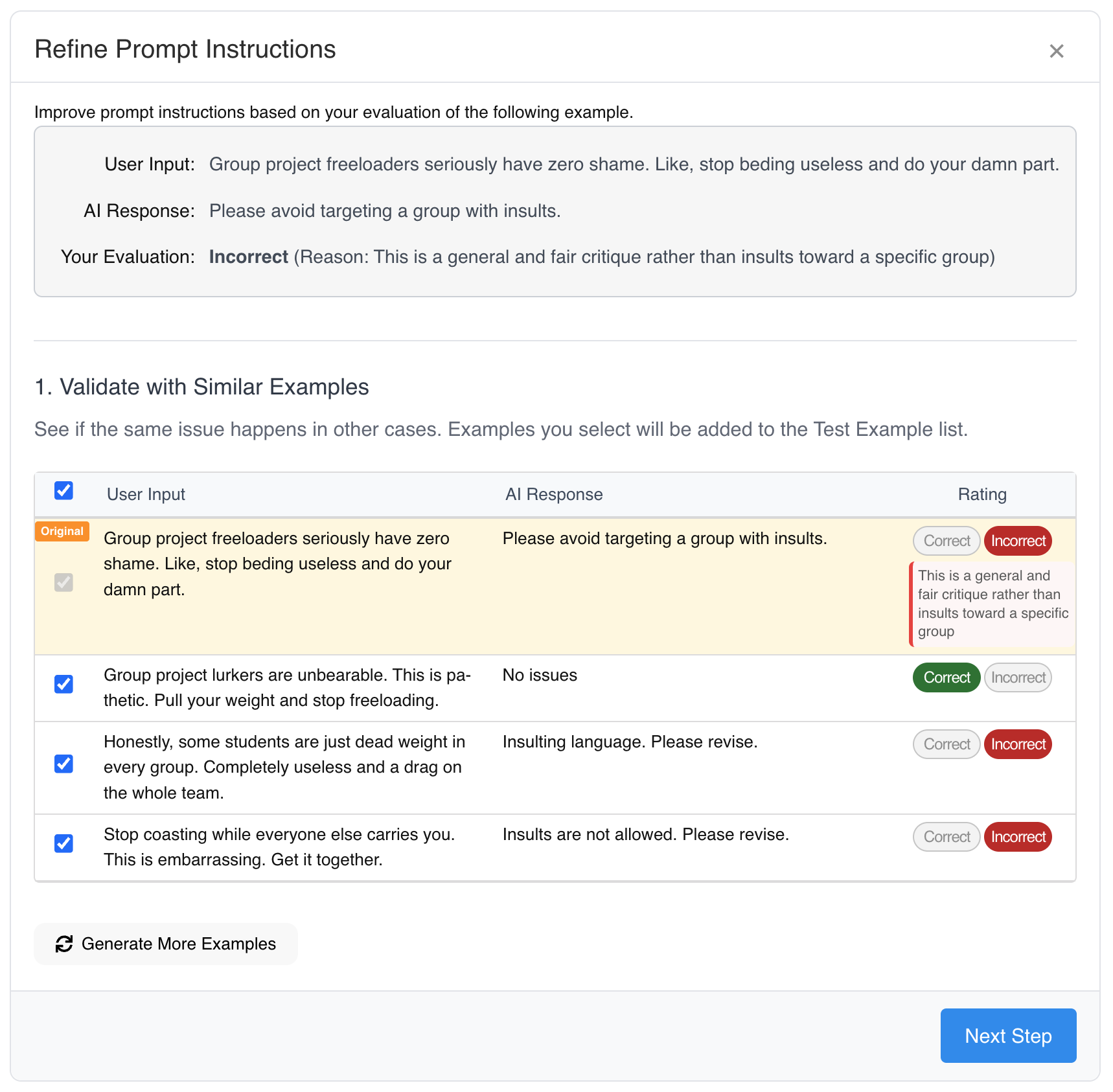}
    \caption{Neighborhood validation step in the workflow. The top section shows the original selected example, along with the user’s label and written rationale. Below, the system provides semantically similar examples that vary slightly from the original, enabling the user to test whether their rationale holds across a neighborhood of cases. As users label these examples, they actively reflect on the policy boundary, and the results directly inform the next step of revising the prompt instruction. The validated examples are then added to the test set, where they continue to be used in subsequent evaluations.}
    \Description{A modal window titled ``Refine Prompt Instructions'' demonstrates the neighborhood probing process. The interface displays a table where the first row, highlighted in yellow and labeled Original, shows a model failure case with a user-provided rationale for an incorrect rating. Below this row, three similar inputs synthesized by the system are listed. Each row includes a user input, the AI Response, and buttons for marking the response as correct or incorrect. Checkboxes on the left allow the user to select specific cases to be added to the test set. At the bottom, buttons labeled ``Generate More Examples'' and ``Next Step'' are displayed.}
    \label{fig:verification}
\end{figure}

\textbf{Step 3: Generalizing the Rationale with Neighborhood Probing.}
The goal of this step is to generalize a single failure into a reusable rule. Our workflow supports \textit{neighborhood probing}: given the user’s rationale, the system synthesizes a set of semantically similar inputs that preserve the same triggers while varying context. 
Clicking a button on the rationale popup opens a larger popup for this purpose. As shown in Figure~\ref{fig:verification}, users can label these cases to assess whether the rationale consistently predicts undesired behavior. 
If the model mishandles all of them, the rationale is confirmed as broadly relevant; if not, users may refine the rationale to be more specific.
These examples will be added to the test set.
We crafted a prompt that generates three new examples as follows:

\begin{lstlisting}
Generate three new user inputs to verify if the failure 
described in the Human Rationale is reproducible, without 
widening scope.

Context
- AI Instruction: {prompt_instruction}
- Original User Input: {user_input}
- AI Response: {ai_response}
- Human Rationale (root cause & semantic triggers): {rationale}

Precedence (for any conflict)
1) Human Rationale (triggers & scope)
2) Original input's style anchors
3) Reference-only info (surface flavor only; ignore if it broadens scope)

Anchors to preserve
- Same semantic trigger(s) as in Human Rationale (paraphrase allowed only if meaning identical)
- Style anchors of the original: language/register, salient markers
...
\end{lstlisting}

\textbf{Step 4: Revising the Specification.}
Once the neighborhood is labeled, the dialog proceeds to the next step: prompt instruction revision. The system synthesizes the original failed example, the user's rationale, and the labels from the neighborhood probe to suggest a revision to the prompt instruction. 
Often, the suggestion takes the form of a new policy line or clarification, such as \textit{``Always answer questions about credit scores, even when they appear in the context of employment, housing, or other life events.''}
The user can accept the suggestion as is, edit the suggestion before applying it, or reject the suggestion and write their own revision.

\textbf{Step 5: Evaluating Against the Test Set.}
Upon applying the revision, several actions are triggered automatically. First, the Prompt Instruction panel is updated with the new version. Second, the original failed example and the additional examples from the neighborhood probe are added to the Test Set panel. Third, the system runs the new prompt instruction against all test cases.

To minimize user effort, an LLM-as-judge automatically provides an initial rating for the new responses. This judge is prompted with the user's previously labeled examples from the test set as few-shot demonstrations, aligning its judgments with the user's specific policies. The user can quickly scan these automated labels and override any that are incorrect. The side-by-side information on the test set table (shown in Figure~\ref{fig:screenshot} top right) clearly shows whether the new prompt instruction fixed the targeted failure. The accuracy metric in the Version History panel updates, providing a quantitative summary of the revision's impact.

\vspace{6pt}

This cycle of discovering a failure, revising the prompt instruction, and evaluating against the test set forms the core loop of data-prompt co-evolution. The user can then return to the Generated Examples panel to probe for the next edge case, continuously strengthening both their test sets and their models in tandem.

\subsection{Implementation Details}

Our system is implemented as a web-based application that runs directly in the browser. The user interface is built with React, while components that require LLM calls are handled by a Python server developed with Flask. The 
server communicates with external LLM APIs to support various stages of the workflow.

We primarily used Gemini 2.5 Flash~\cite{comanici2025gemini} from Google Gemini API\footnote{\url{https://ai.google.dev/gemini-api/docs/models}} to power the workflow steps, such as generating synthetic examples, retrieving similar examples, and suggesting prompt revisions. To ensure fast interactions, reasoning-heavy ``thinking mode'' was avoided in most cases. For model responses to end users, where prompt instructions are simply combined with user inputs, we employed lighter models such as Gemini 2.5 Flash Lite.

Because the LLM-as-judge module runs at every iteration, its runtime must remain low as the test set grows. To ensure responsiveness, we parallelize these LLM calls. A small scalability experiment shows a near-linear speedup. Evaluating 50 examples decreased elapsed time from 52.8 seconds to 5.6 seconds with 10 workers (9.4x speedup) and to 3.4 seconds with 20 workers (15.6x speedup). Although latency is bounded by API response time, parallel processing effectively mitigates this bottleneck and keeps interaction real-time even for larger test sets.

\section{Technical Experiments}
\label{sec:technical}

We conducted simulation-based experiments to evaluate the workflow's core components: \textit{failure discovery} and \textit{instruction refinement}. 
The simulation serves two purposes: testing how well the workflow adapts to diverse, application-specific policy definitions, and validating components that rely on probabilistic LLM outputs in a controlled setting. To do this, we created deterministic ``personas'' that act as ground-truth oracles.

\subsection{Setup}

We utilized two content moderation domains from our user study: an online student forum and restaurant customer reviews (see Section \ref{sec:task} for details). To enable deterministic evaluation, we developed a structured taxonomy and a set of personas with fixed policy preferences.

\textbf{Categorization.} We defined a taxonomy of 24 categorical values spanning four dimensions, designed to capture the factors that shape moderation decisions:

\begin{itemize}
\item Severity: emotional intensity of negativity (e.g., neutral, mild, moderate, strong).

\item Target: who or what is being criticized (e.g., individual, group, general public, service).

\item Content category: the primary issue raised (e.g., facilities, food quality, staff behavior).

\item Problematic style: stylistic markers that escalate perceived harm (e.g., profanity, harsh tone, sarcasm, slang).
\end{itemize}
To map user inputs to these categories, we used a classifier LLM (Gemini 2.5 Flash),
which achieved 89\% accuracy against human-labeled ground truth.
Note that this is the only LLM involved in the experiment pipeline; all moderation decisions made by the personas are fully deterministic, allowing us to evaluate without relying on LLM-based judgments.

\textbf{Simulated Personas (Oracles).} 
For each domain, we defined four distinct personas governed by deterministic logic (if-then rules) based on the taxonomy values. These rules produce both a binary label (i.e., whether user inputs are problematic) and a corresponding rationale. To ensure consistency, each persona uses a predefined rationale template that converts the triggered rule into a short, human-readable explanation. This simulates the ``human rationale'' phase of our workflow and provides structured input to the instruction-refinement module.
While prior research has used LLMs to simulate user judgment, we intentionally avoided LLM-based user simulators to prevent the risk of circular validation, given that our workflow already relies on LLMs in other components.

\begin{itemize}
\item Persona A (Civility): 
prioritizes polite wording; rejects any input containing profanity or insults.

\item Persona B (Expressiveness): 
prioritizes free speech; tolerates harsh tone and slang while rejecting extreme hate speech or threats.

\item Persona C (Fairness): 
prioritizes fairness to individuals and groups; tolerates strong criticism of situations but flags context-free slurs, sweeping denigration, or calls for harm.

\item Persona D (Constructiveness): 
emphasizes informational content; accepts strongly worded complaints but flags vague or purely venting.
\end{itemize}

\subsection{Experiment 1: Validity and Coverage of Failure Discovery}

The workflow begins by synthesizing ``hard'' cases. If these cases are trivial or irrelevant, the co-evolution loop cannot meaningfully proceed. 
We therefore evaluated whether the failure discovery module (Step 1 of the workflow) generates cases that are both valid failures and semantically diverse.

\textbf{Protocol.} 
For each persona, we generated five failure candidates using our workflow and compared them against two baselines: (1) \textit{simple generation}, which prompts the model for typical user inputs; and (2) \textit{adversarial generation}, which prompts the model to generate inputs that cause model failure (similar to red-teaming).
We measured the following two metrics, each averaged over 10 runs:
\begin{itemize}
    \item \textit{Validity:} the proportion of generated examples that were judged incorrect according to the persona's pre-defined determination rules.
    \item \textit{Coverage:} the proportion of generated examples that triggered failures in \textit{at least one} of the four personas. This indicates that the example lies in a policy grey area where reasonable policies can disagree, rather than representing a trivial violation that every persona would reject in the same way.
\end{itemize}

\textbf{Results.} 
Our system outperformed both baselines.
For \textit{validity}, our system achieved a failure discovery rate of $0.635$, performing comparably to the \textit{adversarial} baseline ($0.497$) and far exceeding \textit{simple} generation ($0.031$). 
For \textit{coverage}, while the \textit{adversarial} baseline achieved a score of $0.677$, our system achieved $0.917$.
These results suggest that the workflow produces nuanced, realistic edge cases that span multiple interpretations of policy, rather than just exploiting model weaknesses.

\subsection{Experiment 2: Effectiveness of Instruction Refinement}

We next evaluated the instruction refinement module through two questions: (1) Do the workflow components contribute to instruction improvement? and (2) Does the workflow support cumulative improvement across multiple refinement cycles?

\subsubsection{Ablation Study}

\textbf{Protocol.} 
We isolated the impact of the workflow's unique components by comparing our full workflow against a simplified baseline that skips the rationale-generation and neighborhood-verification steps.

\textbf{Results.} 
The results demonstrate the value of these components. Starting from an initial accuracy of 0.403 on the holdout set (20 examples written by real users collected from the web; details in Section \ref{sec:holdout}), the ablated baseline improved performance marginally to 0.428. In contrast, our full workflow reached 0.502 within a single iteration. This suggests that rationale-driven updates generalize more effectively than narrow, case-specific fixes.

\subsubsection{Sequential Iterations}

\textbf{Protocol.} 
We examined whether the workflow fosters continuous improvement by simulating a session with eight consecutive cycles of discovery and refinement. Each configuration (two domains x four personas) was run ten times.

\textbf{Results.} 
Performance improved steadily across iterations. Accuracy increased sharply after the first iteration (from 0.408 to 0.517), followed by consistent incremental gains across subsequent steps: 0.528, 0.554, 0.570, 0.576, and 0.583 (steps 2--6). Performance peaked at 0.591 at step 7 and stabilized at 0.588 by step 8.
While performance improves steadily, it begins to plateau in later stages. This plateau likely reflects the fact that our simulation relies on personas that encode deterministic and well-scoped policies, allowing many straightforward issues to be addressed early. The remaining errors tend to lie near subtle policy boundaries, leaving less room for large accuracy gains. Meanwhile, real users may iteratively develop and adjust their policies over time, a dynamic that is not fully captured by our fixed personas.

\vspace{6pt}
These experiments demonstrate that the system's two core mechanisms\allowbreak---\allowbreak failure discovery and instruction refinement\allowbreak---\allowbreak operate as intended in a controlled setting. At the same time, 
the observed plateau highlights a broader limitation of simulation-based evaluation: real-world policy design is rarely fixed or deterministic, and instead involves iterative sensemaking as human intent evolves.
Thus, it is necessary to examine whether these capabilities translate into practical utility for actual human users.
We therefore turn to the next section to present a user study comparing our workflow against a human baseline to assess whether it more effectively supports human creativity and iteration.

\section{User Study Setup}

To evaluate how our proposed Co-Evolution workflow facilitates the process of iteratively refining behavioral policies for LLMs in application-specific contexts, we conducted a controlled lab study comparing it against a conventional prompt-editing interface.

\subsection{Study Design}
We employed a within-subjects design, with each participant completing a task in two conditions: our \textit{Co-Evolution} workflow and a \textit{baseline} interface. To mitigate potential learning and carryover effects, we used two distinct yet comparable task domains: content moderation for campus community websites and for restaurant reviews. The order of conditions and the assignment of domains to each condition were counterbalanced across participants using a balanced Latin square. To further reduce domain-specific priming, participants completed a tutorial with a third, distinct domain (YouTube comments) before starting each of the two experimental blocks.
The study was approved by our institution's IRB.

\subsection{Task: Refining a Content Moderation Policy}
\label{sec:task}
Participants assumed the role of policy designers for a new online platform (e.g., a campus forum or a review site), acting as project leads responsible for shaping the platform's behavioral policies. Their goal was to craft a prompt instruction that would guide an LLM to moderate user-submitted comments according to a specific, nuanced policy. The desired policy was to allow critical commentary while filtering out harmful content such as personal attacks, identity-based slurs, and serious, unverified allegations. We explicitly encouraged participants to incorporate their own subjective judgments and to use the provided interface to refine their prompt so that the LLM's behavior reflected their intended policy boundary.

\subsection{Conditions: Co-Evolution vs. Baseline}
The \textit{Co-Evolution} workflow condition provided participants with our full system. This included features for generating synthetic test examples, exploring similar examples around a point of failure, and refining prompt instructions based on suggestions from the system.

The \textit{baseline} condition was a controlled, ablated version of our system that represented a standard prompt-editing environment. The user interface is similar, but with its data-centric workflow disabled. Specifically, it included the prompt editing and history panel on the left and the test set panel on the top right side of the interface where participants could manually add potential user inputs and observe model outputs on these manually crafted inputs. However, the baseline did not offer tools to generate test cases with LLMs, explore the neighborhood space around a failure, or use AI-assisted prompt suggestions.

To ensure a fair comparison and mitigate the difficulty of initiating the test set creation process (``cold start''), participants in both conditions were initially provided with an identical set of three pre-populated test examples. This provided all participants with a consistent initial anchor for their refinement process.

\subsection{Participants}
We recruited 16 participants (5 female; mean age 25.3, SD 2.3) from mailing lists for university students in South Korea. We set two inclusion criteria: participants were required to have used a generative AI service (e.g., ChatGPT) at least ten times in the preceding week; and to have completed at least one university-level AI/ML course or have equivalent experience. These criteria ensured participants were familiar with prompt-based interaction and possessed a foundational understanding of AI concepts, reflecting our target users who are not necessarily AI experts but have sufficient background to engage in prompt refinement and test set construction. We deliberately targeted this demographic to reflect the evolving definition of \textit{developers} or \textit{practitioners}. As LLMs make application development more accessible, developers are increasingly individuals who are not necessarily professional software engineers but understand core AI concepts, specifically test sets and evaluation.
All participants were familiar with Korean, and the pre-populated prompt instructions and test examples were provided in Korean.
Each participant received a gift card (KRW 20,000 $\approx$ USD 15) as compensation.

\subsection{Session Protocol}
Each session lasted approximately 55 minutes and was conducted remotely via video conferencing. After providing electronic consent and completing a brief pre-study questionnaire, participants proceeded through the two counterbalanced experimental blocks. Each block began with a short tutorial and Q\&A session. Participants then completed the task for up to 10 minutes. Following the task, they spent approximately 3 minutes labeling a holdout evaluation set (detailed in the next paragraph). Then they filled out a block-specific post-questionnaire consisting of 10 questions with 7-point Likert scales and 3 open-ended questions.

\begin{figure*}[!t]
    \centering
    \includegraphics[width=0.8\linewidth]{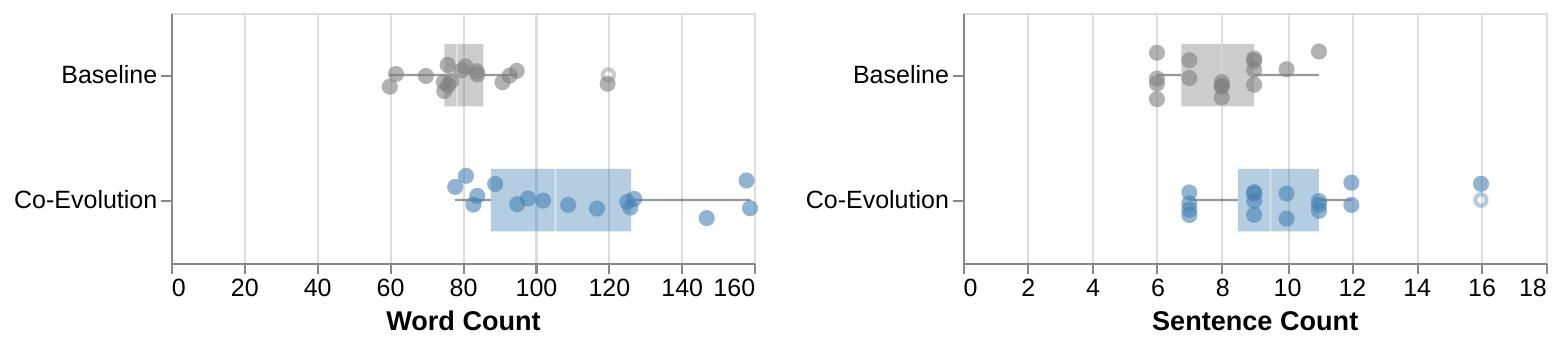}
    \caption{Length of prompt instructions created under each condition, measured in word count and sentence count. Each point represents one participant (jittered for visibility), with box plots summarizing distributions. Instructions produced with the Co-Evolution workflow were significantly longer than those from the baseline condition.}
    \Description{The figure includes two charts, one for word count and one for sentence count. Each chart displays two horizontal box plots, with the baseline condition above and the Co-Evolution condition below, and 16 jittered data points per condition. In both panels, values under the Co-Evolution condition are higher than those under the baseline.}
    \label{fig:wordcount}
\end{figure*}

\begin{figure*}[!t]
    \centering
    \includegraphics[width=1.0\linewidth]{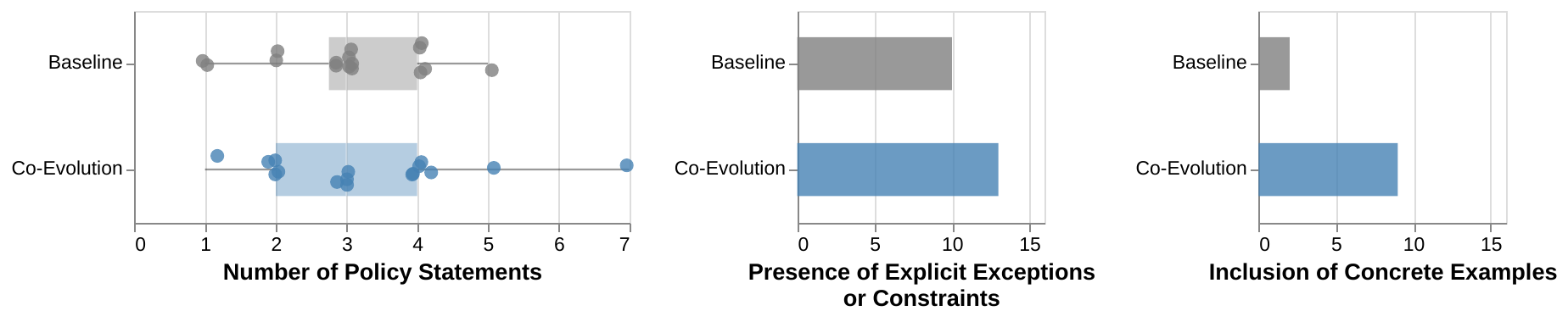}
    \caption{Comparison of how prompt instructions across conditions provided specific information about policy. From left to right: (1) number of policy statements included, (2) whether explicit exceptions or constraints were stated, and (3) whether concrete examples were given. While the number of policy statements was similar, Co-Evolution led participants to add more examples to illustrate.}
    \Description{The figure contains three charts: (1) Number of Policy Statements: this chart shows horizontal box plots, with the baseline condition above and the Co-Evolution condition below. The two conditions have similar medians and upper quartiles, with modest differences in the lower quartile and maximum values. (2) Presence of Explicit Exceptions or Constraints: this chart shows a horizontal bar chart with one bar per condition, indicating that the measured value is slightly higher for the Co-Evolution condition. (3) Inclusion of Concrete Examples: this chart shows a horizontal bar chart with one bar per condition, indicating a substantially higher measured value under the Co-Evolution condition.}
    \label{fig:chart-structual}
\end{figure*}

\subsection{Holdout Evaluation Sets}
\label{sec:holdout}
To quantitatively measure the model behavior for the prompt instructions created by participants, we constructed two domain-specific \textit{holdout evaluation sets}.
These sets were distinct from the test sets created within the Co-Evolution interface and were used solely for post-task evaluation.

Each holdout evaluation set was composed of 20 items we collected from real-world online communities: a private forum website for university students in South Korea and a local restaurant review website. 
To construct a challenging yet balanced test bed, we curated the items using stratified sampling. Each set is composed of
8 \textit{borderline} items (nuanced cases designed to be ambiguous and probe the subjective decision boundary of a policy) and
12 \textit{clear-cut} items (less ambiguous cases, further split into 9 clearly \textit{no violation} and 3 clearly \textit{violation} instances).
This stratified composition, balancing clear-cut cases with ambiguous borderline ones, is a common practice for building robust evaluation benchmarks in the field of NLP~\cite{rajpurkar-etal-2018-know, kwiatkowski-etal-2019-natural, nie-etal-2020-adversarial}. It was also informed by our formative interviews, which highlighted the need to evaluate policies against examples with a range of difficulty.

At the end of each experiment block, participants labeled the above 20 items in the corresponding holdout set as either acceptable or problematic. These labels served as a personal \textit{ground truth} reflecting their intended policy (for RQ 4). We then applied each participant’s refined prompt (from baseline and Co-Evolution conditions) to the same 20 items using the Gemini 2.5 Flash Lite model.

\begin{figure*}[!t]
    \centering
    \includegraphics[width=0.8\linewidth]{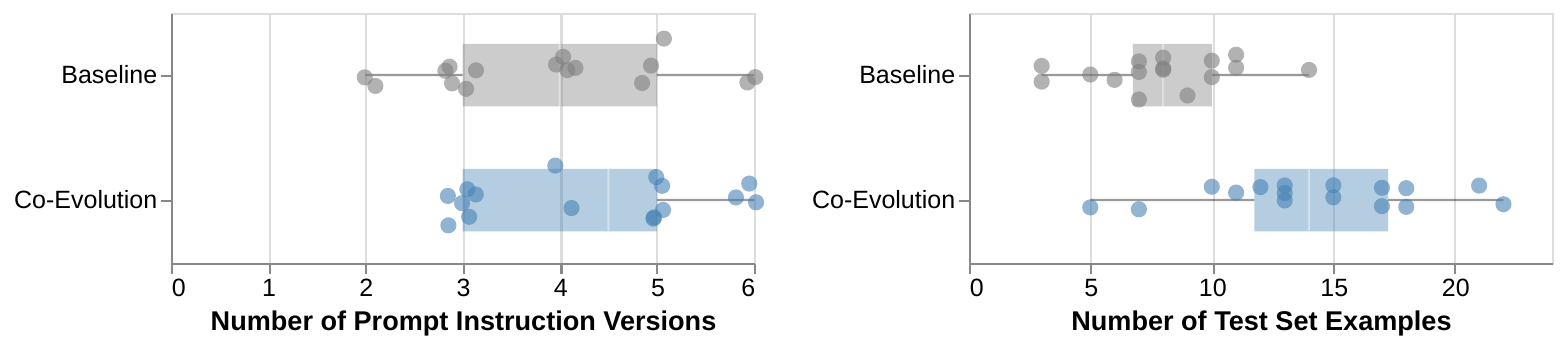}
    \caption{Prompt refinement process measured by the number of instruction versions (left) and the test set size (right). 
Participants produced a similar number of versions across conditions, and those using Co-Evolution generated substantially larger test sets than in the baseline condition.}
    \Description{The figure includes two box-plot panels comparing conditions. One panel shows the number of prompt instruction versions, and the other shows the number of test set examples. In both panels, each condition is displayed as a horizontal box plot, with the baseline condition above and the Co-Evolution condition below. For prompt instruction versions, the two conditions show similar distributions, with similar medians and interquartile ranges. For test set examples, the Co-Evolution condition shows higher values than the baseline.}
    \label{fig:chart-iterations}
\end{figure*}

\section{Study Results}

\subsection{RQ 1: How do prompt instructions refined by our workflow differ from the baseline?}

We first compared prompt instructions created with our Co-Evolution workflow against the baseline method along two dimensions: length (how long the instructions were) and specificity (how detailed and clear they were).

\paragraph{Length.} Prompt instructions created with the Co-Evolution workflow were significantly longer than those from the baseline.
The average word count was 111.1, compared to 81.2 in the baseline condition. 
A paired $t$-test confirmed this difference was statistically significant ($p<0.001$).
This trend was consistent across both task domains (campus: 107.8 vs. 78.9 words; reviews: 114.5 vs. 83.5 words).
Sentence counts were also higher under Co-Evolution (9.8 vs 7.9),
and a paired $t$-test indicated that the difference is significant ($p=0.019$).

\paragraph{Specificity.}
To measure specificity, we assessed three structural features of the prompt instructions:
(1) the number of \textit{policy statements}, i.e., individual sentences articulating specific rules or criteria about the model behavior;
(2) the presence of \textit{explicit exceptions or constraints}, such as ``do not'' rule or boundaries of the task; and
(3) the inclusion of \textit{concrete examples} that illustrate policies.

The results are shown in Figure~\ref{fig:chart-structual}.
The average number of \textit{policy statements} 
did not differ significantly between the two conditions
(i.e., 3.00 for the baseline vs. 3.31 for Co-Evolution).
However, prompt instructions from the Co-Evolution workflow more frequently included \textit{explicit exceptions or constraints} (13 vs. 9 out of 16 participants) and \textit{concrete examples} (9 vs. only 2). 
These results suggest that our workflow helps participants articulate boundaries and illustrate rules with examples, producing instructions that are not only longer but also richer in detail.

\subsection{RQ 2: How does the workflow shape the prompt refinement process?}

We next examined how the process of reaching the final prompt instructions varied across conditions. We analyzed participants' effort and engagement by measuring the number of instruction revisions and the size of the test sets.

Participants produced a comparable number of revisions in the two conditions. 
On average, they saved 3.88 versions in the baseline and 4.31 versions in Co-Evolution, a difference that was not statistically significant (distributions shown in Figure~\ref{fig:chart-iterations}).

However, a significant difference was observed in the size of the test sets created, as expected thanks to the neighborhood probing feature. Participants using the Co-Evolution workflow generated substantially larger test sets (avg. 14.19 items) compared to those in the baseline (avg. 7.94 items). 
A $t$-test confirmed this difference was highly significant ($p < 0.001$).

The similarity in revision counts is noteworthy. While one might assume the simpler baseline interface would lead to faster or more frequent iterations, our findings suggest that the cognitive load of manually imagining test examples may have balanced out this simplicity. In contrast, the Co-Evolution workflow, by guiding and partially automating test case generation, enabled participants to build more comprehensive test examples within a similar number of revision cycles, fostering a more structured refinement process.

\subsection{RQ 3: Do instruction revisions lead to changes in model behavior?}

We then asked whether revisions to prompt instructions translated into actual changes in model outputs. To examine this, we compared model responses between the initial default instruction provided in the interface and the final instructions written by participants. 
For each of the 20 questions in the \textit{holdout evaluation set}, we compared whether the model’s output under the default prompt matched that under the participant’s final revised prompt.

\begin{figure}[!tb]
    \centering
    \includegraphics[width=1.0\linewidth]{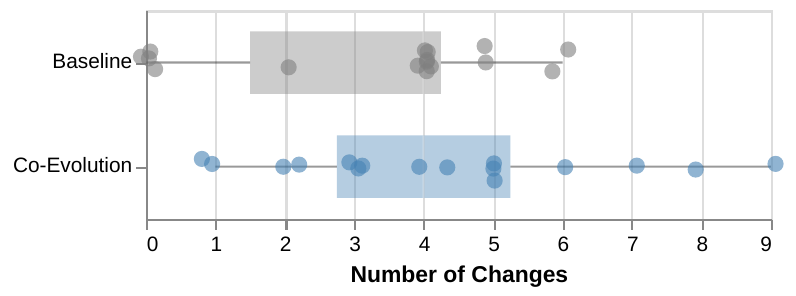}
    \caption{Number of output changes per participant under the two conditions. 
Each point shows an individual participant (jittered for visibility), with box plots summarizing the distributions. 
In the baseline condition, four participants produced no changes; Co-Evolution yielded a more consistently distributed pattern across participants.}
    \Description{This chart shows paired horizontal box plots comparing the number of changes across conditions. The baseline condition includes four observations at zero, with the remaining values extending up to six. In contrast, the Co-Evolution condition exhibits a more evenly distributed range of values, spanning from one to nine.}
    \label{fig:chart-changes}
\end{figure}

In the baseline condition, participants produced an average of 3.25 outcome changes relative to the default prompt instruction. 
In the Co-Evolution condition, the average was higher at 4.25 changes.
The difference in counts was not statistically significant.  
The shape of the distribution, however, revealed contrasts (shown in Figure~\ref{fig:chart-changes}).
In the baseline condition, four participants produced no changes at all, 
and the Shapiro-Wilk test indicated a non-normal distribution.
In contrast, in the Co-Evolution condition, every participant induced at least one change, and the distribution satisfied normality.

\subsection{RQ 4: Does the workflow improve alignment between instructions and model behavior?}

Our next research question asked whether prompt instructions revised through our workflow led to model behavior that better matched participants’ intentions. As described in the previous subsection, participants labeled 20 potential user inputs, and we compared these labels against the model's response.

\begin{figure}[!t]
    \centering
    \includegraphics[width=1.0\linewidth]{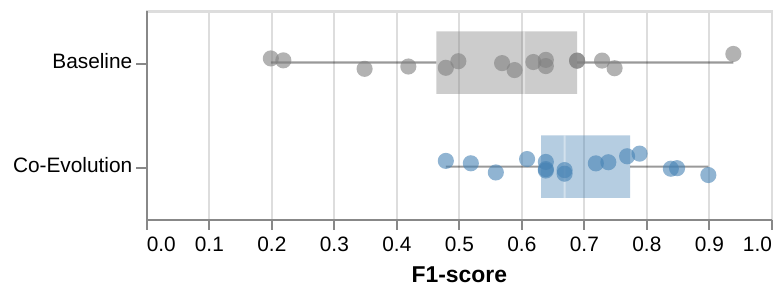}
    \caption{Alignment between model outputs and participant labels, measured by F1-score. 
Participants’ revised instructions in Co-Evolution yielded higher alignment than in the baseline. 
A paired $t$-test indicated this difference was significant at the 95\% confidence level ($p=0.0241$).}
    \Description{This chart shows paired horizontal box plots comparing F1-scores across conditions. The Co-Evolution condition has a higher median and higher first and third quartiles than the baseline. While the maximum F1-score is slightly higher in the baseline condition, the Co-Evolution condition shows a narrower overall range, with a substantially higher minimum value.}
    \label{fig:chart-alignment}
\end{figure}

While the baseline condition achieved an average F1-score of 0.56, our Co-Evolution workflow achieved 0.69 (distributions shown in Figure~\ref{fig:chart-alignment}).
To assess whether this improvement was statistically reliable, we first conducted a Shapiro-Wilk test of normality, which indicated no violation ($p = 0.33$). A paired $t$-test then showed that the improvement was significant at the 95\% confidence level ($p = 0.0241$).

These results highlight that the instruction revisions observed in RQ 3 were not arbitrary: they shifted the model's behavior closer to participants' intended judgments. They suggest that the Co-Evolution workflow can guide participants toward revisions that both alter model behavior and bring it into better alignment with their goals.

To rule out the possibility that these results were influenced by task-order learning effects, we performed additional analyses. First, to examine whether participants improved simply by performing one task before the other, for each participant, we computed the accuracy difference between the second and first task. Performance on the first task was slightly higher on average, and the difference was not statistically significant ($p=0.31$).
We also compared the advantage of the Co-Evolution condition across interface orderings (participants who began with the baseline vs. those who began with Co-Evolution). Again, we found no significant differences (Welch's $t$-test $p=0.74$), indicating that the observed benefit of Co-Evolution is unlikely to result from learning effects.

\subsection{RQ 5: What are the efficacy and coverage of the test sets created?}
To investigate the quality of the constructed test sets, we analyzed the data along two dimensions: \textbf{efficacy} (measuring how effectively examples exposed model limitations) and \textbf{coverage} (measuring the semantic diversity of the test scenarios).

First, to assess the \textit{efficacy} of the test sets, we examined whether the generated examples actively probed model limitations rather than being trivial inputs. We measured the rate at which test inputs successfully triggered a model failure---specifically, examples participants explicitly marked as incorrect at least once during the session. The Co-Evolution workflow demonstrated a significant advantage: 51.9\% of the test examples (93 out of 179) were flagged as failures, whereas the baseline condition yielded only 26.7\% (23 out of 86). This indicates that the Co-Evolution workflow more effectively guides users to generate challenging edge cases that probe the model's weaknesses.

Second, to determine the \textit{coverage} of the test sets, we examined the semantic diversity of inputs to ensure they addressed a broad range of scenarios. Specifically, for each participant, we calculated the coverage based on the four categories (e.g., target group, problematic style) defined in our technical experiment (Section \ref{sec:technical}). The Co-Evolution condition achieved a significant higher coverage of 58\% (avg. 13.8 out of 24 categories), compared to the baseline, which covered 49\% (avg. 11.7 categories). 
While this increase is partly attributable to the higher total number of examples generated in the Co-Evolution condition, the result suggests that the workflow successfully scaffolded participants in broadening their data space to cover a more diverse range of scenarios, rather than simply generating more of the same types of errors.

\subsection{RQ 6: How do participants perceive the effectiveness of the workflow?}

The above five research questions analyzed what changed in the prompt instructions and how those changes affected model behavior. We now shift from these outcome-focused measures to participants’ own subjective assessments.

\begin{figure}[!t]
    \centering
    \includegraphics[width=1.0\linewidth]{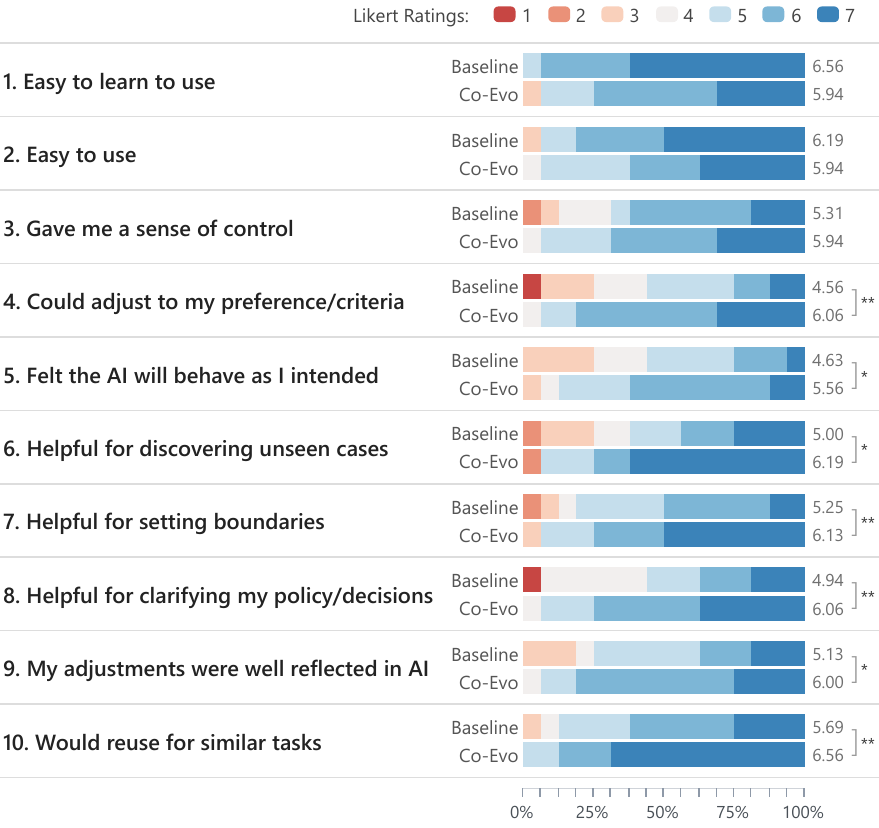}
    \caption{Subjective ratings for 10 questionnaire items, comparing the baseline and Co-Evolution (Co-Evo) conditions. Bars represent the distribution of responses on a 7-point Likert scale, with darker shades indicating stronger agreement. Average scores are shown to the right of each bar, and significance markers denote differences between conditions, with * indicating $p < 0.05$ and ** indicating $p < 0.01$.}
    \Description{This figure shows ten questionnaire items presented as paired horizontal percentage stacked bar charts, with the baseline condition above and the Co-Evolution condition below. Each bar represents the distribution of responses on a 7-point Likert scale. Mean scores are displayed to the right of each chart, and asterisks indicate statistically significant differences between the conditions.}
    \label{fig:ratings}
\end{figure}

Figure~\ref{fig:ratings} shows the distribution of subjective ratings for the 10 questions for each of the two conditions.
First, both conditions scored highly on basic usability. The baseline, with fewer features, was rated slightly higher on ease of learning, as expected (Q1), and both conditions were judged similarly easy to use 
(Q2; 5.94 vs. 6.19). Even on sense of control (Q3), where we anticipated the baseline might be stronger, ratings were comparable, with Co-Evolution even a bit higher (mean: 5.94 vs. 5.31). These patterns suggest that the two systems were broadly on par for straightforward ease of use.

Where the Co-Evolution workflow stood out was in the dimensions it was designed to support. Participants reported that it helped adapt the system to their own preferences and criteria, with 13 of 16 gave ratings of 6 (agree) or 7 (strongly agree) compared to 4 under the baseline (Q4; $p < 0.01$).
They also reported greater confidence that the AI would behave as intended (Q5; $p < 0.05$) and found the workflow more effective for covering the ``long tail'', such as discovering unseen cases (Q6; $p < 0.05$) and defining policy boundaries in realistic scenarios (Q7; $p < 0.01$). Participants also emphasized that the workflow clarified their policy decisions where 12 vs. 6 participants rated at 6 or 7 (Q8; $p < 0.01$). Furthermore, participants perceived that their adjustments were better reflected by the AI using our workflow (Q9; 13 vs. 6 rated at 6 or 7; $p < 0.05$), echoing the stronger alignment performance shown in RQ 4. Lastly, participants were more satisfied overall, with stronger intentions to reuse it for future tasks (Q10; $p < 0.01$).

Together, these ratings indicate that participants perceived clear added value from the Co-Evolution workflow, beyond the basic level of usability that both systems achieved.

\section{Discussion}

Based on the qualitative observations and participant feedback from the user study, we discuss how our interactive system facilitates the iterative co-evolution process and reflect on its limitations, which suggest design implications for future systems.

\subsection{Feedback on Strengths}

\subsubsection{Enabling Fine-Grained Specification through Concrete Examples}
A significant advantage reported by participants was the ability to define nuanced, domain-specific policies by directly interacting with concrete data examples. This was particularly crucial for ambiguous tasks like content moderation, where abstract guidelines often fail to capture real-world complexities. By reviewing and labeling diverse examples, participants could more easily identify and articulate the subtle boundaries of their criteria. One participant noted how this process helped them refine their standards:
\textit{``By creating a diverse test set for `Incorrect' items where the boundary was ambiguous, I was able to consider more detailed criteria.''}

This hands-on approach, grounded in tangible data, transformed the abstract task of policy-making into a more manageable and effective process. Another participant confirmed this, stating:
\textit{``Seeing diverse expressions [of criticism] that I could encounter in real life with my own eyes was very helpful in setting more detailed standards for the acceptable level of criticism.''}
This aligns with Policy Maps~\cite{lam2025policy}, which showed that surveying concrete cases helps designers define custom regions in the space of model behavior.

\subsubsection{Facilitating Nuanced and Rapid Co-evolution}

Participants highly valued the system's interactivity. The rapid feedback loop allowed them to quickly test hypotheses and iterate on their prompts, making the refinement process more efficient and intuitive. One participant remarked:
\textit{``It was impressive that it processed feedback in real-time and let me modify the prompt command again.''}

The comparison table view was essential for helping users understand the concrete impact of their changes. By visualizing the outputs of the old and new prompt instructions simultaneously, users could easily assess whether their modifications led to the desired improvements. This feature directly supported the goal-oriented nature of their task.
\textit{``Being able to compare the results of the old prompt and the new prompt at a glance was helpful in writing the prompt's content (e.g., community guidelines) because I could see what difference the new prompt made.''}
This underscores the importance of not just enabling rapid iteration but also providing tools that make the outcomes of those iterations transparent and understandable.

\subsubsection{Scaffolding Human Judgment with Just-in-Time AI Assistance}

One of our design philosophy was to use AI to augment humans. The study revealed that this approach was effective. Participants found AI-generated suggestions, such as rationales for an incorrect judgment, to be particularly helpful for articulating their own reasoning, a task that can be cognitively demanding.
\textit{``When I rated the AI's evaluation as `Incorrect', it was helpful that it suggested reasons why. The reason for choosing `Incorrect' can be ambiguous to put into words, but seeing the options suggested by the AI is helpful.''}

By offloading ancillary tasks like verbalizing a rationale, the system allowed users to focus their cognitive resources on the core challenge: making nuanced judgments. This form of ``just-in-time'' assistance demonstrates a promising direction for human-AI collaboration.

\subsection{Limitations and Design Opportunities}
Despite its strengths, our study also highlighted limitations that offer valuable opportunities for future research.

\subsubsection{Beyond Binary Judgments: The Need for Handling Ambiguity}
The system's reliance on a binary correct/incorrect classification can be a limitation. While simple, this binary choice forced participants into difficult decisions for borderline cases, which are often the most crucial for refining a model's understanding of nuance. A participant described this struggle:
\textit{``There are clear criteria for inappropriate posts, but the boundary isn't sharp, which made evaluation feel difficult. And since there were only two choices, `Correct' and `Incorrect,' for the AI's judgment, I was conflicted on how to judge ambiguous, in-between examples.''}

This is a well-known challenge in interactive machine learning~\cite{dudley2018review}. As suggested by prior work~\cite{kulesza2014structured, chang2017revolt}, systems might provide users with ways to express uncertainty, such as adding a ``Maybe'' category. Such inputs could trigger additional components, such as seeking further clarification from the user, or explicitly modeling this uncertainty.

\subsubsection{The Challenge of Diverse Example Generation for Creative Exploration}
While the example generation feature was intended to help users discover edge cases and stimulate creative exploration (akin to red teaming), participants noted a lack of diversity in the generated outputs:
\textit{``When generating user examples, it was a bit inconvenient that the examples generated at once were not very diverse and seemed similar.''}

This highlights a persistent challenge in generative AI: ensuring the novelty and breadth of generated content~\cite{whitney2024real}. Although our system included mechanisms to encourage variety, more sophisticated strategies are needed. Future work could explore techniques for further promoting semantic diversity in generated examples, allowing users to guide the generation process toward specific types of edge cases to broaden the scope of exploration.

\subsubsection{Managing Policy and Test Set Complexity}
While our workflow helps refine policies by starting from a single failing example, it does not fully address policy conflicts in complex applications. For instance, as users continuously add examples, the test set itself becomes a new source of complexity, presenting a significant scaling challenge. One participant pointed this out:
\textit{``The structure inevitably leads to an expanding test set over time. For stability this is good, but I think it would be better to be able to hide sets that have shown no change in response.''}

This highlights a critical design need: managing the cognitive load associated with a large, living test set. Future systems must therefore not only help resolve policy conflicts but also provide scalable interfaces for managing the test data itself. This could involve workflows for synthesizing insights from a cluster of related failures, advanced filtering or summarization mechanisms to help users focus, and tools for visualizing how different subsets of the test data are affected by a change in policy~\cite{cabrera2023zeno, kahng2024llm, sivaraman2025divisi}.

\subsubsection{Navigating the Double-Edged Sword of Automated Judgments}
The use of an LLM-as-judge~\cite{zheng2023judging} presents a compelling but complex trade-off. On one hand, it significantly reduces the burden of manual labeling, as discussed earlier. On the other hand, its fallibility can mislead users or foster over-reliance. The same participant who praised the AI's assistance also voiced a critical concern:
\textit{``After improvement, the Correct/Incorrect predictions were often correct, but there were occasional periods where judgment became clouded. I'm concerned that some users might give up on thinking for themselves and just automate it.''}

This captures the double-edged nature of AI assistance. An incorrect judgment from the AI can send a user down the wrong path, while consistently correct judgments may lead to automation bias, eroding the user's critical engagement. This implies that systems employing AI judges must be designed with caution. Future interfaces might consider explicitly communicating the AI's confidence levels, and designing workflows that actively encourage, or even require, human oversight in the loop.

\subsubsection{Extending to Open-Ended Generative Tasks.}

While our user study focused on binary content moderation tasks to ensure quantifiable evaluation, the proposed workflow is conceptually applicable to open-ended generative tasks too (e.g., summarization or creative writing). If a model's generated response is unsatisfactory, the user can still identify the failure and articulate a rationale to guide the refinement. However, we note that applying our system to generative contexts may require more sophisticated feedback mechanisms beyond binary labeling. Future work should investigate additional interaction designs to support these scenarios effectively, for instance, by integrating user-defined rubrics~\cite{kim2024evallm} or example demonstrations~\cite{petridis2024constitutionmaker}.

\subsubsection{Scaling to Collaborative Teams.}
Our study focused on a foundational scenario of a single application developer refining an LLM's behavior via its prompt instruction. 
This setup reflects an increasingly common situation, especially in one-person startups. 
However, real-world applications often scale to a collaborative team of product managers, policy experts, and engineers.
This shifting from a solo developer to a team
introduces significant CSCW (Computer-Supported Cooperative Work) challenges around creating shared test sets, resolving disagreements, and others~\cite{rastogi2024insights, subramonyam2021towards, sivaraman2025tempo}.
Future research should address how the co-evolution workflow adapts to these dynamics, ensuring that the feedback loop remains tight.

\subsubsection{From Prompting to Model Parameter Updates.}
While our work focuses on refining prompts, which is the primary interface available to application developers today, the idea of co-evolution can be extended more broadly as a relationship between data and models. In this view, beyond updating prompt instructions, one could aim to fine-tune underlying LLMs by updating their parameters (e.g., model weights). However, unlike prompts, which can be revised rapidly, updating model parameters poses greater technical challenges. Encouragingly, recent advances in parameter-efficient fine-tuning (PEFT), including low-rank adaptation (LoRA), have begun to reduce this latency~\cite{li2021prefix, lester2021power, hu2022lora}. As these techniques mature, the co-evolution approach explored in our work could potentially be applied to updating model weights, increasing its practical value in settings where prompt-based control is impractical or impossible, such as under strict privacy constraints.

\section{Conclusion}

In this work, we introduced \textit{data-prompt co-evolution}, a new workflow where prompt instructions and test data are treated as co-evolving artifacts rather than separate, static resources. Through the design of an interactive workflow and its evaluation in a controlled user study, we showed that this approach enables application developers to refine nuanced policies more concretely, construct richer test sets systematically, and achieve more consistent alignment between their intentions and model behavior than with a baseline prompt-editing tool. Our findings suggest that embedding iterative, evidence-grounded test-set growth directly into the development loop can democratize LLM refinement, moving beyond ad-hoc tinkering or episodic red-teaming toward continuous, human-in-the-loop improvement. We highlighted future opportunities for scaling co-evolution to larger test sets, collaborative settings, and model weight updates, advancing the design of responsible and human-centered AI systems.

\begin{acks}
We thank the members of the Human-Data Interaction Lab at Yonsei University for their feedback. This work was supported in part by the Yonsei University Research Fund (2025-22-0155) and an IITP grant funded by MSIT, Korea (RS-2024-00353131).
\end{acks}

\bibliographystyle{ACM-Reference-Format}
\bibliography{references}

\end{document}